\documentstyle[aps,prd,epsfig]{revtex}
\newcommand{\bee}{\begin{equation}}
\newcommand{\ee}{\end{equation}}
\newcommand{\beea}{\begin{eqnarray}}
\newcommand{\eea}{\end{eqnarray}}

\preprint{COLO-HEP-458
}
\begin{document}

\title{Comment on ``Evidence Against Instanton Dominance of Topological
 Charge Fluctuations in QCD'' by I. Horv\'ath {\it et al.}}
\author{Thomas DeGrand, Anna Hasenfratz}
%
%
\address{
Department of Physics,
University of Colorado, 
        Boulder, CO 80309 USA
}
\date{\today}
\maketitle
\begin{abstract}
We comment on the recent paper (hep-lat/0102003) by Horv\'ath, Isgur, McCune,
and Thacker, which concludes
 that the local chiral structure of fermionic eigenmodes
is not consistent with instanton dominance. Our calculations, done with
an overlap action, suggest the opposite conclusion.
\end{abstract}
\pacs{11.15.Ha, 12.38.Gc, 12.38.Aw}
%
%

\section{Introduction}
In a recent paper, Ref. \cite{ref:them},
Horv\'ath, Isgur, McCune, and Thacker present evidence from lattice simulations
claiming to show that the behavior of fermionic eigenmodes in the QCD
vacuum is inconsistent with an instanton picture.
They refer to a recent paper of ours, Ref. \cite{ref:us}, and remark
``Although their qualitative conclusions regarding the instantons differ
 from ours, there does not seem to be any direct conflict
 between their lattice results and ours.'' We disagree with 
that statement. To expose that disagreement, we reanalyzed  a subset
of our data along the lines of Ref. \cite{ref:them}.

Why is it important to resolve this  disagreement? There is an apparent
 inconsistency,
first described by Witten in 1979 \cite{ref:witten79}, between large
number of colors (large-$N_c$) dynamics and instanton-based phenomenology.
We do not wish to summarize arguments favoring one or the other approach,
which are well-described in a variety of publications \cite{ref:largeN}.
In principle, lattice simulations might distinguish between the qualitative
features of the two phenomenologies. However, the lattice calculations of
Refs. \cite{ref:them} and \cite{ref:us} produce contradictory results.

The two simulations differ in several ways. Both are done in quenched
approximation, using the Wilson gauge action. Ref. \cite{ref:them} works
at a lattice coupling $\beta=5.7$ and Ref. \cite{ref:us} works at
$\beta=5.9$. The lattice spacings $a$ are at nominal values of 0.17 and 0.11
fm, respectively,  from the Sommer parameter,
 using the interpolating formula of Ref. \cite{ref:precis}.
The fermion action of Ref. \cite{ref:them} is the Wilson fermion action.
This action has rather poor chiral properties, as evidenced by its
large additive quark mass renormalization and the presence of exceptional
configurations. Formally, it has $O(a)$ scaling violations. Its real
eigenmodes are not eigenfunctions of $\gamma_5$, and the expectation value
of $\gamma_5$ of a real eigenmode is a function of the eigenvalue.

The action of Ref. \cite{ref:us} is an overlap action \cite{ref:neuberfer},
which is constructed by using an approximate overlap action as its
starting point\cite{ref:TOM_OVER}.  This action has exact chiral symmetry.
There is no additive mass renormalization of the quark mass and
there are no exceptional
configurations. Zero-eigenvalue eigenmodes are chiral eigenstates.
Nonzero eigenvalue eigenmodes have zero expectation value of $\gamma_5$.
The action has only $O(a^2)$ discretization artifacts.

Our action uses ``fat links.''
Fat link actions replace the usual one-link gauge connection with a
 combination of several gauge paths. This combination might or might
 not be projected to SU(3), but in any case it is gauge invariant
 and local. Locality assures that the fat link action is in  the same
 universality class as the thin link one.
Fat links improve the chiral behavior of nonchiral actions \cite{ref:fat},
and their use in our implementation of the overlap action 
is simply to make the calculation of the overlap operator more
efficient (a factor of roughly 20 for the action of Ref. \cite{ref:TOM_OVER},
compared to the use of the thin link Wilson action in the overlap.)
The particular choice of fattening is not too important.
In Ref. \cite{ref:us}, we used an APE-blocked
link \cite{ref:APEblock} as the gauge connection. 
In Ref. \cite{ref:Bernard} it is shown that the perturbative effect
of N levels of APE-smearing with a smearing parameter $\alpha$ is
to multiply the vertices by a form factor $(1 - \alpha {\hat q}^2 /6)^N$
with $q$ the gluon momentum and $\hat q^2 = 4/a\sum_\mu \sin^2 q_\mu a/2$.
In coordinate space this corresponds to a Gaussian smearing with a spread
$\langle x^2 \rangle = \alpha N/3 a^2$ of the quark gluon vertex.
Some readers might be
concerned that several levels of 
APE smearing might adversely affect our results.
To address that point, we describe results from another fattening, one
chosen to lie rigorously within a hypercube \cite{ref:HYP_Thermo}.
We observe no qualitative change in our results.

\section{Analysis of Fermion Eigenmodes}

The statement of the problem has been given by Ref. \cite{ref:them}:
``If an instanton-dominated picture of low eigenmodes is at all
valid, we would expect the peaks of the wave function to closely
 resemble instanton
zero modes. Thus, if instantons dominate, a local peak in the wave function
for a low-lying eigenmode (a fermion lump) should be dominantly
a lump in $\psi_L^\dagger\psi_L$ or $\psi_R^\dagger\psi_R$, but not both.''
(Here $\psi_L^\dagger\psi_L = \psi^\dagger (1-\gamma_5)\psi $ and
 $\psi_R^\dagger\psi_R = \psi^\dagger(1+\gamma_5)\psi $.)
``On the other hand, fermion lumps without a definite chirality
 would be an indication of non-self-dual fluctuations, as
 expected from a confinement-related mechanism.''

In Ref. \cite{ref:us}we used an overlap action with $N=10$ levels of APE smearing with $\alpha=0.45$ APE parameter. We measured the autocorrelation function
of the local chirality density
 $\omega(x) = \langle \psi(x) | \gamma_5 | \psi(x) \rangle$ 
and the correlator of $\omega(x)$ with a pure gauge observable sensitive to
topological charge $Q(x)$. We saw that both
 kinds of correlators
were strongly peaked at small operator separation indicating the the chiral
 density has spatially localized lumps and those lumps correlate with the
 lumps of the topological charge density. However, there is
a logical possibility that a lump in 
$\omega(x)=\bar\psi_L\psi_L-\bar\psi_R\psi_R$  does not indicate a lump in
 either $\bar\psi_L\psi_L$ or $\bar\psi_R\psi_R$.
To check
that possibility, we follow Ref. \cite{ref:them} and define a ``chiral
 order parameter'' $X(x)$ 
\bee
\tan({\pi \over 4}(1 + X(x)) = {{|\psi_L(x)|}\over{|\psi_R(x)|}}
 = ( {{\psi_L^\dagger(x)\psi_L(x)}\over{\psi_R^\dagger(x)\psi_R(x)}} )^{1/2}
\ee
and plot histograms of the density of the variable $X$, $\rho(X)$,
for regions of the lattice where $\psi(x)^\dagger\psi(x)$ is large,
a few per cent of the points of the lattice. We distinguish
two cases:

(a) Zero modes are chiral. The analogs of Figs. 1 and 4 of Ref. \cite{ref:them}
for our action are just delta functions $\rho(X) \simeq \delta(X \pm 1)$.
The chiral symmetry breaking inherent in the Wilson fermion action of
Ref. \cite{ref:them} broadens this distribution.

\begin{figure}[thb]
\begin{center}
\epsfxsize=0.45 \hsize
\epsffile{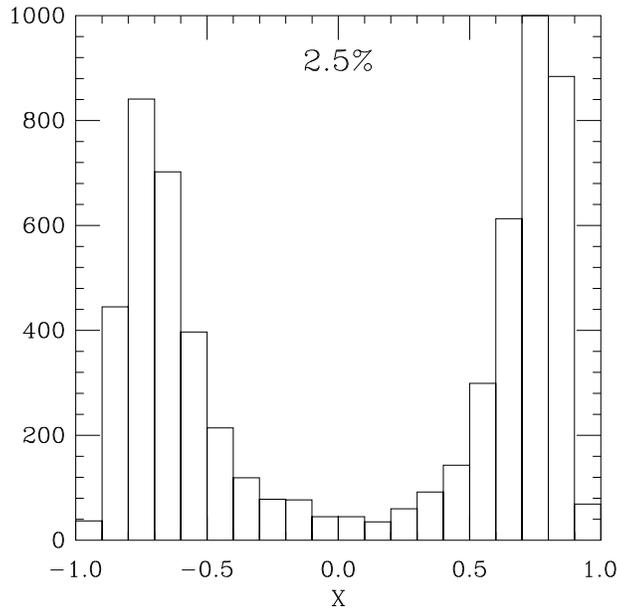}
\end{center}
\caption{
Histogram of $\rho(X)$ of the lowest two nonchiral eigenmodes for
the overlap action, with a cut keeping the top 2.5 per cent of
$\psi^\dagger\psi$.
}
\label{fig:hl2a}
\end{figure}

(b) Nonzero eigenvalue eigenmodes are nonchiral, but we find that the
 distribution $\rho(X)$ is strongly peaked near $X = \pm 1$.
To show this, we take a set of six $12^4$ lattices from our
data set and plot histograms of $\rho(X)$.
 Fig. \ref{fig:hl2a} shows the result for the lowest
 two nonchiral eigenmodes for
the overlap action, with a cut keeping the top 2.5 per cent of
$\psi^\dagger\psi$. The imaginary part of the eigenvalue of the Dirac operator
for these modes is in the range $0.03/a$ to $0.09/a$. Keeping more lattice points in the histogram will fill in the dip. That simply implies that as we keep more and more lattice points we
 pick up more of the vacuum fluctuations around the chiral lumps and
 those eventually overwhelm the chiral structure. This happens very slowly with the two smallest non-chiral modes. Even with 30 per cent of the lattice points the two peak structure is clearly visible.

\begin{figure}[thb]
\begin{center}
\epsfxsize=0.45 \hsize
\epsffile{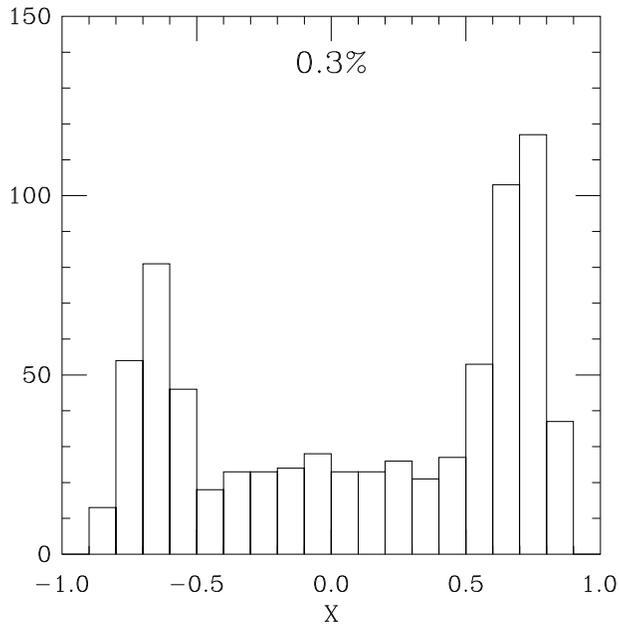}
\end{center}
\caption{
Histogram of $\rho(X)$ of the highest two eigenmodes (out of ten) for
the overlap action, with a cut keeping the top 0.3 per cent of
$\psi^\dagger\psi$.
}
\label{fig:hh205a}
\end{figure}

\begin{figure}[thb]
\begin{center}
\epsfxsize=0.45 \hsize
\epsffile{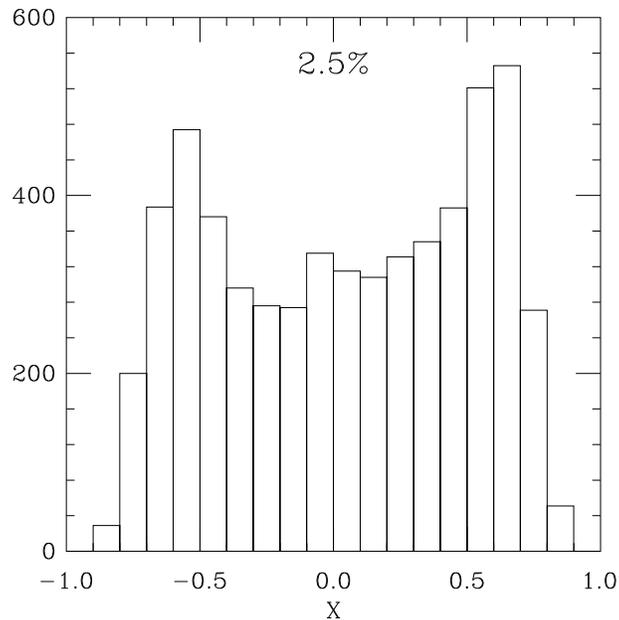}
\end{center}
\caption{
Histogram of $\rho(X)$ of the  highest two eigenmodes (out of ten) for
the overlap action, with a cut keeping the top 2.5 per cent of
$\psi^\dagger\psi$.
}
\label{fig:hh23a}
\end{figure}

 Figs. \ref{fig:hh205a} and  \ref{fig:hh23a}
 shows the result for the largest modes we have recorded,
the ninth and tenth eigenvalue modes of the squared Dirac operator.
The imaginary part of the eigenvalue of the Dirac operator for these modes
is around $0.20/a$-$0.26/a$.
Figure \ref{fig:hh205a}, where about the top 0.3 per cent
 of $\psi^\dagger\psi$ is
kept in the histogram, shows strong peaks around $X=\pm 1$.
 The peaks
disappear as we add more points to the distribution. With 2.5 per cent of the
lattice points kept the histogram is almost flat as Figure \ref{fig:hh23a}
 shows.

Lower eigenvalue eigenmodes show stronger peaking than higher eigenvalue
modes, but all the modes we examined show peaking. We remind the reader that
we showed in Ref. \cite{ref:us} that hadronic correlators for light
 quark masses are well approximated by quark propagators containing only a
few eigenmodes.

Next we consider results obtained with an overlap action that uses 
 more local fat links.
The fat links of the hypercubic blocking \cite{ref:HYP_Thermo}
mix links only within a hypercube  but achieve
 almost the same level of smoothness as the APE smearing
 considered previously.

\begin{figure}[thb]
\begin{center}
\epsfxsize=0.45 \hsize
\epsffile{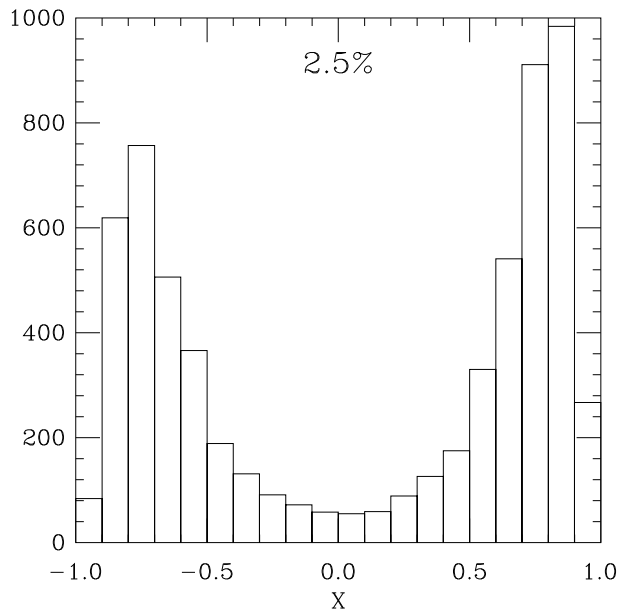}
\end{center}
\caption{
Histogram of $\rho(X)$ of the lowest two nonchiral eigenmodes (out of ten) for
the overlap action, with a hypercubic fat link,
 with a cut keeping the top 2.5 per cent of
$\psi^\dagger\psi$.
}
\label{fig:hhl2a}
\end{figure}

The analog of Fig. \ref{fig:hl2a} from eigenmodes of the overlap
action with hypercubic fat link is shown in Fig. \ref{fig:hhl2a}.
We used the same $12^4$ configurations and kept the same fraction of lattice points as in Fig. \ref{fig:hl2a}.
The shape of the curve is qualitatively similar to that from APE-blocked
links. Peaks of fermion density are again chiral.

\section{Conclusions}
Our 
studies with a chiral fermion action at a lattice spacing near 0.11 fm showed 
that low eigenvalue fermionic eigenmodes have a structure strongly correlated
with the locations of instantons and anti-instantons. This note demonstrates
that in the locations where the fermionic modes are largest, the modes
are also chiral.
It seems to us that the simplest description of what we see is in terms of
an instanton liquid model of the QCD vacuum.

We do not doubt that the numerical results of Ref. \cite{ref:them} are correct,
but we are concerned that the combination of a larger lattice spacing
and the use of a nonchiral fermion action introduces lattice artifacts
which compromise extrapolation to the continuum limit. 

\section*{Acknowledgments}
We acknowledge conversations with I.~Horvath and H.~Thacker
about the points we have raised in this paper.
 This work was supported by the
U.~S. Department of Energy.


\end{document}